\documentclass[preprint,3p,times,onecolumn]{elsarticle}
\usepackage{amsmath,amssymb,pennames,fixltx2e,lineno}

\usepackage{hyperref}

\def\d{\mathrm{d}}
\def\p{\partial}

\renewcommand{\vec}[1]{{\bf #1}}
\newcommand{\mat}[1]{#1}

\newcommand*\patchAmsMathEnvironmentForLineno[1]{%
  \expandafter\let\csname old#1\expandafter\endcsname\csname #1\endcsname
  \expandafter\let\csname oldend#1\expandafter\endcsname\csname end#1\endcsname
  \renewenvironment{#1}%
     {\linenomath\csname old#1\endcsname}%
     {\csname oldend#1\endcsname\endlinenomath}}%
\newcommand*\patchBothAmsMathEnvironmentsForLineno[1]{%
  \patchAmsMathEnvironmentForLineno{#1}%
  \patchAmsMathEnvironmentForLineno{#1*}}%
\AtBeginDocument{%
\patchBothAmsMathEnvironmentsForLineno{equation}%
\patchBothAmsMathEnvironmentsForLineno{align}%
\patchBothAmsMathEnvironmentsForLineno{flalign}%
\patchBothAmsMathEnvironmentsForLineno{alignat}%
\patchBothAmsMathEnvironmentsForLineno{gather}%
\patchBothAmsMathEnvironmentsForLineno{multline}%
}

\begin{document}

\begin{frontmatter}

\title{Optimized differential energy loss estimation for tracker detectors}

\author{Ferenc Sikl\'er}
\ead{sikler@rmki.kfki.hu}
\address{KFKI Research Institute for Particle and
Nuclear Physics, Budapest, Hungary \\ CERN, Geneva, Switzerland}

\author{S\'andor Szeles}
\address{E\"otv\"os University, Budapest, Hungary}

\begin{abstract}

The estimation of differential energy loss for charged particles in tracker
detectors is studied. The robust truncated mean method can be generalized to
the linear combination of the energy deposit measurements. The optimized
weights in case of arithmetic and geometric means are obtained using a detailed
simulation. The results show better particle separation power for both
semiconductor and gaseous detectors.

\end{abstract}


\end{frontmatter}


\section{Introduction}

The identification of charged particles is crucial in several fields of
particle and nuclear physics: particle spectra, correlations, selection of
daughters of resonance decays and for reducing the background of rare physics
processes \cite{Ullaland:2003cm,Yamamoto:1999uc}.
Tracker detectors, both semiconductor and gaseous, can be employed for particle
identification, or yield extraction in the statistical sense, by proper use of
energy deposit measurements along the trajectory of the particle.
While for gaseous detectors a wide momentum range is available, in
semiconductors there is practically no logarithmic rise of differential energy loss
($\d E/\d x$) at high momentum, thus only momenta below the the minimum ionization region are
accessible.
In this work two representative materials, silicon and neon are studied.
Energy loss of charged particles inside matter is a complicated process. For
detailed theoretical model and several comparisons to measured data see
Refs.~\cite{Bichsel:1988if,Bichsel:1990}.

While the energy lost and deposited differ, they will be used interchangeably
in the discussion. It is also clear that the energy read out also varies due to
noise and digitization effects.

This article is organized as follows: Sec.~\ref{sec:simulation} describes the
microscopical energy loss simulation used in this study.
Sec.~\ref{sec:truncated} introduces the basic method of truncated mean, while
Sec.~\ref{sec:weighted} deals at length with the optimization of weighted
arithmetic and geometric means. Possible handling of different path lengths
is discussed in Sec.~\ref{sec:pathLength}. Results of the simulation and
applications of the optimized weighted means are shown in
Sec.~\ref{sec:results}. The work ends with conclusions and it is supplemented
by four Appendices with interesting results, such as optimal weights in case of
few (App.~\ref{sec:few}) and many measurements (App.~\ref{sec:many}); some
theoretical insights (App.~\ref{sec:equally}); also on connection to maximum
likelihood estimation (App.~\ref{sec:connection}).

\begin{figure*}

 \begin{center}
  \input{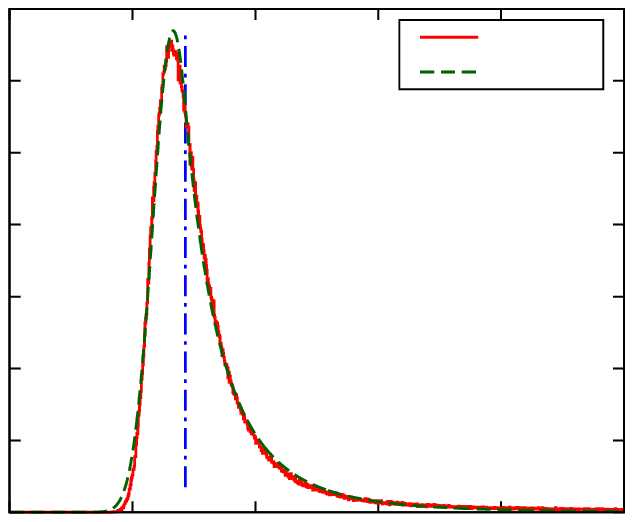}
  \input{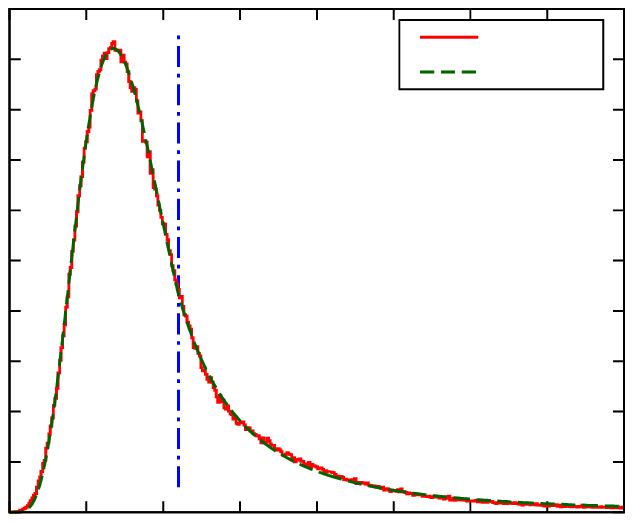}
 \end{center}

 \caption{Comparison of differential energy loss distributions for 300~$\mu$m
silicon (left) and 1~cm neon (right), at $\beta\gamma = 3.17$. The probability
density function (solid) is shown with theory motivated fits (dashed). Above a
certain $\d E/\d x$ value, indicated by the vertical dash-dotted lines, a power
function was used, while below that the product of a power and a Gaussian
(silicon) or the product of a power and an exponential (neon) was taken. For
details see App.~\ref{sec:equally}.}

 \label{fig:distro}

\end{figure*}

\section{Simulation}
\label{sec:simulation}

When a charged particle traverses material it loses energy in several discrete
steps, dominantly by resonance excitations ($\delta$-function) and Coulomb
excitations (truncated power-law term). This latter is the reason for the long
tail observed in energy deposit distributions.

The probability of an excitation, energy deposit, along the path of the
incoming particle is a function of $\beta\gamma = p/m$ of the particle and
depends on properties of the traversed material. The conditional probability
density $p(\Delta|t)$, deposit $\Delta$ along a given path length $t$, can be
built using the above mentioned elementary excitations combined with an
exponential occurrence model. The details of the microscopical simulation can
be found in Refs.~\cite{Bichsel:1988if}, \cite{Bichsel:1990} and
\cite{Bichsel:2006cs}.  The result of these recursive convolutions is a smooth
asymmetric density distribution with long tails (see Fig.~\ref{fig:distro},
solid lines). In order to model detector and readout noise, Gaussian
random values with standard deviation of 2~keV (0.01~keV) were added to each
hit for silicon (neon). For further studies on noise dependence see
Sec.~\ref{sec:other}.

\section{Truncated mean}
\label{sec:truncated}

There are several possibilities for the estimation of the differential energy loss
of a charged particle. An approach using some theoretical model of energy loss
would enable to use advanced methods such as maximum likelihood estimation.
However, especially at startup, particle detectors are not expected to be
understood to the degree that would enable the use of such estimator. The
results would be quite sensitive to the choice of the model, precision of
detector gain calibration, the level of noise
and several backgrounds.

One of the robust and simple estimators is the so called truncated mean
that is traditionally used in gas filled detector chambers
\cite{Fernow:1986,Grupen:2008}. It reduces the influence of
high energy deposits in the tail of the energy deposit
distribution. For this aim a given fraction of the upper 30-60\% (and
sometimes the lower 0-10\%) measurements are discarded and only the remaining
measurements are averaged with equal weights.
Recent studies show that five or even four layers of silicon allow to reach
10\% resolution, using the truncated mean method \cite{Yamamoto:1999uc}.

Let $\Delta_i$ denote the deposit and $t_i$ denote the path length in the active
material of the detector, in case of the $i$th measurement of the particle
trajectory.
The differential energy loss is $y_i \equiv \Delta_i/t_i$. The numbering of the
measurements is such that they are ordered: $y_i \le y_{i+1}$, in case of $n$
space points ($i = 1, 2, \dots, n$). The estimator is simply
\begin{gather*}
 y = \frac{\sum_{i=1}^n w_i y_i}
          {\sum_{i=1}^n w_i}.
\intertext{E.g. if only the
lower half of the hits are used, (0\%,50\%) truncation, the corresponding weights $w_i$ are}
 w_i = \begin{cases}
       0   & \text{if $2i > n+1$} \\
       1/2 & \text{if $2i = n+1$} \\
       1   & \text{if $2i < n+1$}.
       \end{cases} 
\end{gather*}

In the rest of this paper by {\it simple truncated mean} we will mean this
(0\%,50\%) truncation.

\begin{figure*}
 
 \begin{center}
  \input{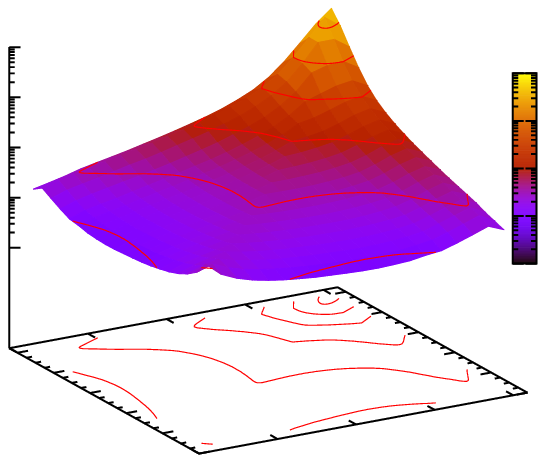}
  \input{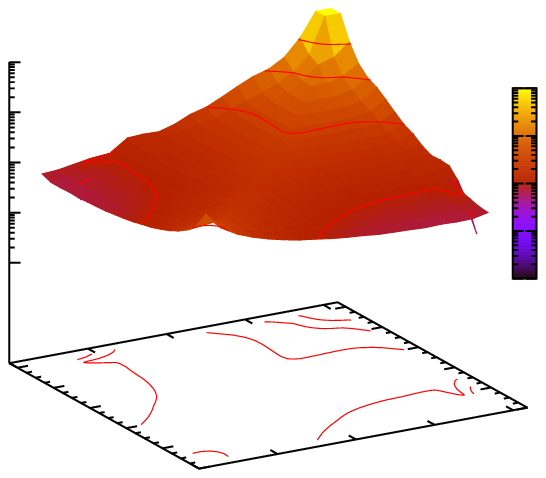}
 \end{center}

 \caption{Correlation matrix of hits $(i,j)$ for 300~$\mu$m silicon (left) and
1~cm neon (right), at $\beta\gamma = 3.17$, in case of 20 hits on track.}

 \label{fig:correlation}

\end{figure*}

\section{Weighted means}
\label{sec:weighted}

It is possible to generalize this estimator, and optimize the weights, by
looking at some measures of its distribution. The
generalization can be twofold. Instead of a simple truncated mean, the more
general weighted mean, linear combination, can be examined where the constant
weights are allowed to take on different values, not just 0, 1/2 or 1. In addition
it is possible that the performance of the weighted mean is more beneficial
when averaging a monotonic function of the measurements $x_i = R(y_i)$ than
just taking the $y_i$ values themselves. It is clear that the transformed
values are also ordered: $x_i \le x_{i+1}$. Let us look at the linear
combination of $n$ measurements
\begin{align}
 y &= R^{-1} \left(\sum_{i=1}^n w_i R(y_i)\right), &
 R(y) \equiv x &=  \sum_{i=1}^n w_i x_i \label{eq:estimator}
\end{align}

\noindent where $\sum_{i=1}^n w_i = 1$. In the following two cases will be
examined further. If $R$ is identity we get back the weighted arithmetic mean,
the use of $R(y) = \log y$ gives the weighted geometric mean. While the former
choice is historical and most simple, the geometric mean has its root in the
behavior of energy loss distribution since that can be approximated by
log-normal distribution. In that sense $\log(y)$ seems to have a more symmetric
distribution than the long tailed $y$. 

In fact an optimization of the transformation function $R$ could complete the
study, but that task appears to be highly non-trivial. Both arithmetic and
geometric means are special cases among power functions, $R(y) = y^p$, with
$p=1$ and $p\rightarrow 0$, respectively:
\begin{align*}
 y_\text{arith} &= \sum w_i y_i, &
 y_\text{geom}  &= \exp \left( \sum w_i \ln y_i \right).
\end{align*}

\noindent Note that extreme cases $p = -\infty$ (minimum) and $p = \infty$
(maximum) are also included in the weighted arithmetic mean.

\subsection{Goal of the optimization}

In order to facilitate clean particle identification and extraction of particle
yields, the distribution of the estimator should be narrow and should be
sensitive to changes in the average energy loss. If the momentum of the particle or
its mass is altered the distribution of deposited energy will change
accordingly. A small change can be modelled by multiplying each energy deposit
along the trajectory with a factor $1+\alpha$, where $\alpha$ is small. If the
distribution of the estimator is close to Gaussian (mean $m$, standard
deviation $\sigma$), its mean
will shift. The separation power between the original and altered distributions, in
units of standard deviation, will be
\begin{equation*}
 \frac{1}{\sigma} \frac{\p m}{\p \alpha}.
\end{equation*}

\noindent In case of the arithmetic mean, the estimator is linear, $\p m/\p
\alpha = m$, thus the relative resolution $\sigma/m$ has to be minimized (see
Sec.~\ref{sec:arith}). For the geometric mean the multiplication
corresponds to a shift $\alpha$, hence the the absolute resolution $\sigma$
is to be minimized (see Sec.~\ref{sec:geom}).

The mean of the $i$th ordered measurement and the covariance of the $i$th and
$j$th measurements play a central role in the optimization.  They are
determined as
\begin{align*}
 m_i    &= \langle x_i \rangle, &
 V_{ij} &= \langle x_i x_j \rangle - \langle x_i \rangle \langle x_j \rangle.
\end{align*}

\noindent Both $m_i$ and $V_{ij}$ can be estimated using detailed physics
simulation described in Sec.~\ref{sec:simulation}. The correlation matrix for
silicon and neon is shown in Fig.~\ref{fig:correlation}, for a given thickness
and $\beta\gamma$ choice, in case of 20 hits on track. Higher
deposits are strongly correlated, thus they contain less information and
should get less weight than other hits. The mean and variance
of the estimator $\sum_{i=1}^n w_i x_i$ (Eq.~\eqref{eq:estimator}) are
\begin{align}
 m        &= \sum_{i=1}^n m_i w_i, &
 \sigma^2 &= \sum_{i,j=1}^n w_i V_{ij} w_j.
 \label{eq:moments}
\end{align}

With help of optimized weights not only the differential energy loss, but also its
variance can be estimated, a definite advantage over the simple truncated or
other plain averages. The weights are of practical use if, for a wide range of
number of hits on track, they appear independent of or insensitive to
$\beta\gamma$ values and material thickness. In the following it is shown that
this is indeed the case.

\subsection{Weighted arithmetic mean}
\label{sec:arith}

The task is to minimize the relative resolution
\begin{equation*}
 \frac{\sigma}{m} =
   \frac{\sqrt{\vec{w}^T \mat{V} \vec{w}}}{\vec{m} \vec{w}}
\end{equation*}

\noindent by varying the weights $\vec{w}$. (Note that here we switched to
vector and matrix forms.) The square of this quantity is
\begin{equation*}
 q(\vec{w}) = \frac{\vec{w}^T V \vec{w}}{(\vec{w}^T \vec{m}) (\vec{m}^T \vec{w})} =
\frac{\vec{w}^T V \vec{w}}{\vec{w}^T M \vec{w}}.
\end{equation*}

\noindent The term on the right side is a generalized Rayleigh quotient, $M =
\vec{m} \otimes \vec{m}^T$ is a dyadic matrix.  For the first variation $\delta
q$ 
\begin{gather*}
  \delta q = \frac{\delta[\vec{w}^T (V - q  M) \vec{w}]}{\vec{w}^T M \vec{w}} \\
\intertext{A vector $\vec{w}$ that minimizes $q$ must give $\delta q = 0$ and thus
satisfy $V \vec{w} = q M \vec{w}$,
which can be rearranged to get}
 \vec{w} = q (V^{-1} \vec{m}) (\vec{m}^T \vec{w}).
\end{gather*}

\noindent Here the vectors $\vec{w}$ and $V^{-1} \vec{m}$ should be parallel.
If sum of the weights has to be one, the optimal weights are
\begin{equation}
 \vec{w} = \frac{V^{-1} \vec{m}}{\vec{1}^T V^{-1} \vec{m}}
 \label{eq:arithweight}
\end{equation}

\noindent where $\vec{1}$ is a column vector of ones. It follows that
the value of the relative resolution at the minimum is
\begin{equation*}
 \min\left(\frac{\sigma}{m}\right) =
  \sqrt{q} = \frac{1}{\sqrt{\vec{m}^T V^{-1} \vec{m}}}.
\end{equation*}

The sensitivity on the weights could be obtained from the Hessian

\begin{equation*}
 H = 2 \frac{V - q M}{(\vec{w}^T \vec{m}) (\vec{m}^T \vec{w})}.
\end{equation*}

\noindent Since at minimum $H \vec{w} = \vec{0}$, according to Cramer's rule
$\operatorname{det} H = 0$, so $H$ is singular.  The sensitivity on weights,
corresponding to 1\% increase of the relative resolution, can be
demonstrated as
\begin{equation}
 \Delta w_i = \sqrt{10^{-2} \cdot 2 q / H_{ii}}.
 \label{eq:sensitivity}
\end{equation}

\begin{figure}

 \begin{center}
  \input{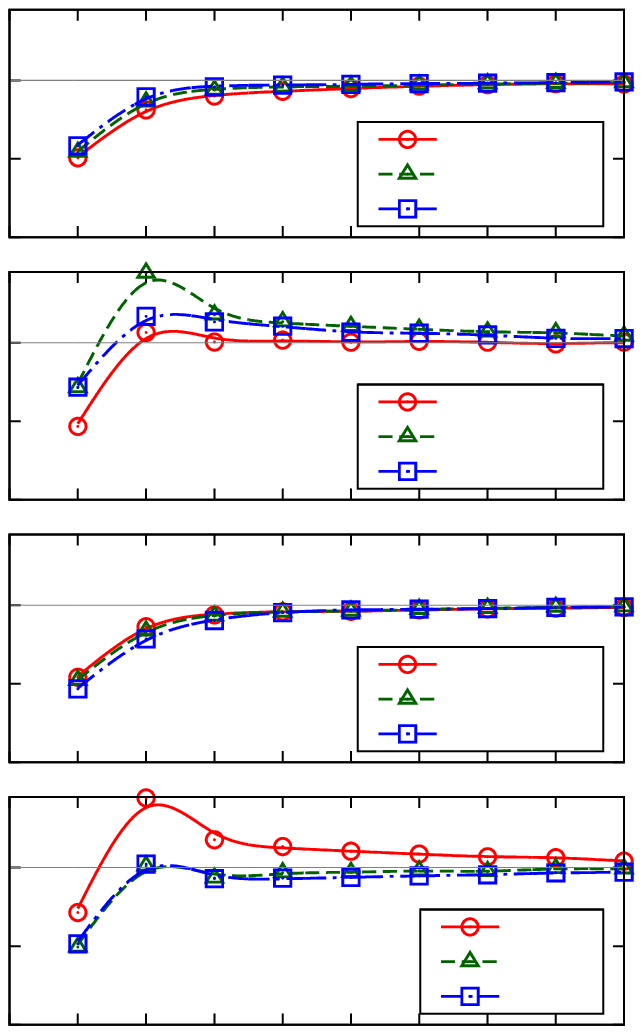}
 \end{center}

 \caption{From top to bottom: normalization factors for 300~$\mu$m silicon and
1~cm neon, at various $\beta\gamma$ values; normalization factors in case of
$\beta\gamma = 3.17$ for silicon and neon, at various thicknesses. Lines are
drawn to guide the eye.}

 \label{fig:normalization}

\end{figure}

\subsection{Weighted geometric mean}
\label{sec:geom}

The task is to minimize the variance $\sigma^2$, with the constraint $\sum_i
w_i = 1$. Thus, the expression to minimize is
\begin{equation*}
 q(\vec{w}) =
   \vec{w}^T \mat{V} \vec{w} + \lambda \left(\vec{1}^T \vec{w} - 1\right).
\end{equation*}

\begin{figure*}

 \begin{center}
  \input{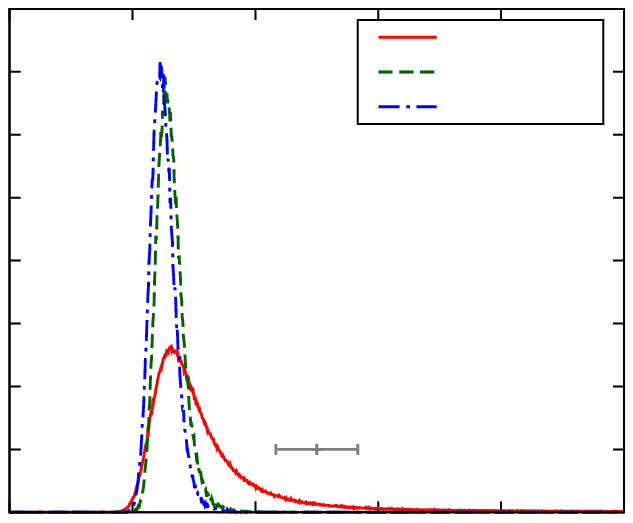}
  \input{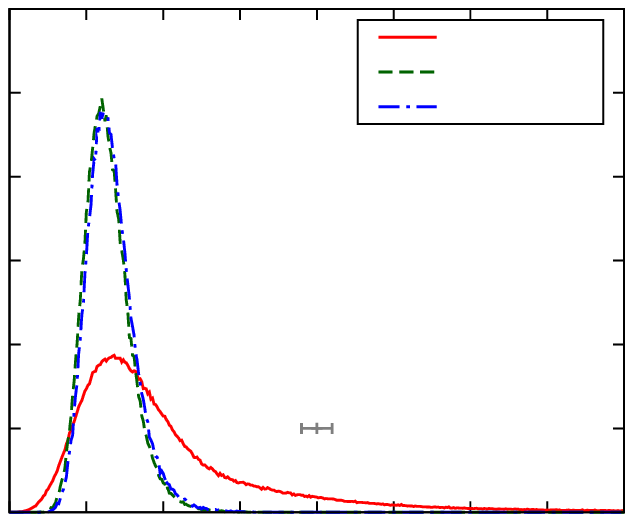}
 \end{center}

 \caption{Comparison of differential energy loss distributions for 300~$\mu$
silicon (left) and 1~cm neon (right), at $\beta\gamma = 3.17$. The probability
density function (solid) is shown with estimators obtained using the simple
truncated mean (dashed) and the weighted arithmetic mean after optimization
(dash-dotted), in case of 6 hits on track. The horizontal bar indicates the
magnitude of noise (2~keV for silicon, 0.01~keV for neon), multiplied by a
factor 10.}

 \label{fig:estimators}

\end{figure*}

\noindent where $\lambda$ is a Lagrange multiplier. At the minimum there is
linear system of equations to solve
\begin{align*}
 2V \vec{w} + \lambda \vec{1} &= \vec{0}, &
 \vec{1}^T \vec{w} - 1 &= 0 
\end{align*}

\noindent or in matrix form
\begin{align*}
 H
 \begin{pmatrix} 
 \vec{w} \\ \lambda
 \end{pmatrix}
 &= 
 \begin{pmatrix} 
 \vec{0} \\ 1
 \end{pmatrix},
 &
 H &= 
 \begin{pmatrix} 
   2V & \vec{1} \\
   \vec{1}^T & 0
 \end{pmatrix}
\end{align*}

\noindent
where $H$ is also the Hessian of $q$. The block matrix $H$ can be inverted
\begin{equation*}
 H^{-1} = \frac{1}{\vec{1}^T V^{-1} \vec{1}}
 \begin{pmatrix}
  - V^{-1} - V^{-1} \vec{1} \vec{1}^T V^{-1} &  V^{-1} \vec{1} \\
  \vec{1}^T V^{-1} & -2
 \end{pmatrix}
\end{equation*}

\noindent and the equations are solved. The optimal weights are
\begin{equation}
 \vec{w} = \frac{V^{-1} \vec{1}}
      {\vec{1}^T V^{-1} \vec{1}}
 \label{eq:geoweight}
\end{equation}

\noindent while the multiplier is $\lambda = - 2/(\vec{1}^T V^{-1} \vec{1})$.
A comparison of Eq.~\eqref{eq:arithweight} and Eq.~\eqref{eq:geoweight} shows
that both expressions have a similar structure. It follows that the value of
the standard deviation at the minimum is
\begin{equation*}
 \min(\sigma) = \frac{1}{\sqrt{\vec{1}^T V^{-1} \vec{1}}}
\end{equation*}

\noindent which is at the same time the relative resolution of the
back-transformed $\exp(x) \equiv y$. The sensitivity on weights,
corresponding to 1\% increase of the relative resolution can be obtained.

\subsection{Rescaling the weights}

Note that the resulted weights for both the arithmetic and geometric means are
functions of the number of measurements $n$. Similarly, both the mean and the
variance of the estimators vary with $n$. In order to eliminate the dependence
of the mean, the weights are renormalized by taking the $n~\rightarrow~\infty$
limit as reference
\begin{equation*}
 \vec{w'}(n) = \vec{w}(n) \cdot \frac{m(\infty)}{m(n)}.
\end{equation*}

Normalization factors for the arithmetic mean are shown in Fig.~\ref{fig:normalization}. It is clear
the values quickly converge to 1 as the number of hits on track increases. Even
at low hit values ($n \le 4$) the factors are quite independent of
$\beta\gamma$ and thickness, this is especially true for silicon.

\begin{figure*}[!h]

 \begin{center}
  \input{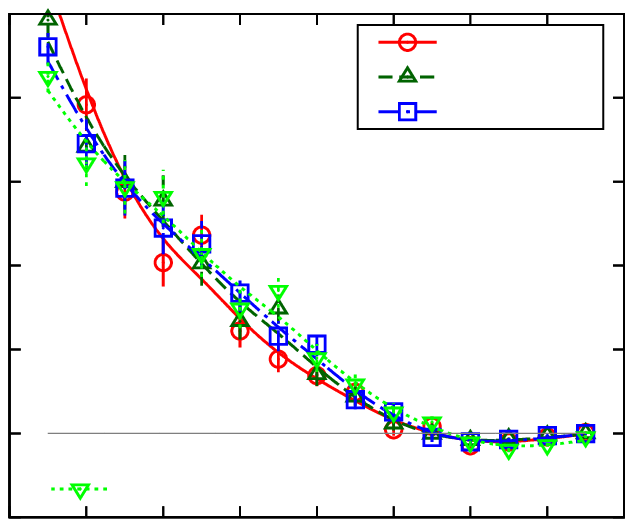}
  \input{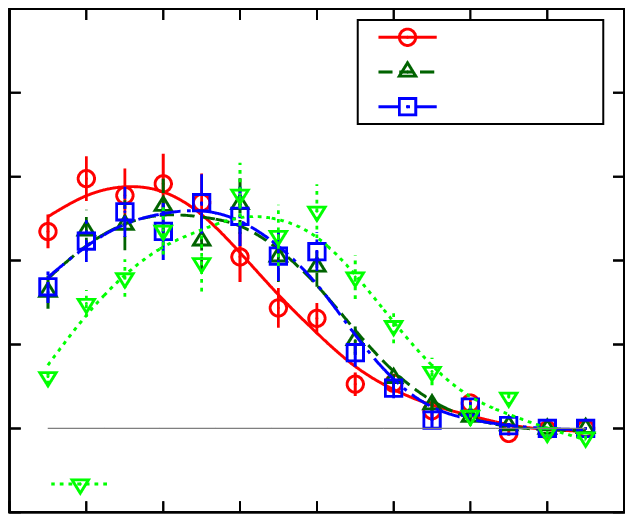}
 \end{center}
 
 \vspace{-0.3in}
 
 \begin{center}
  \input{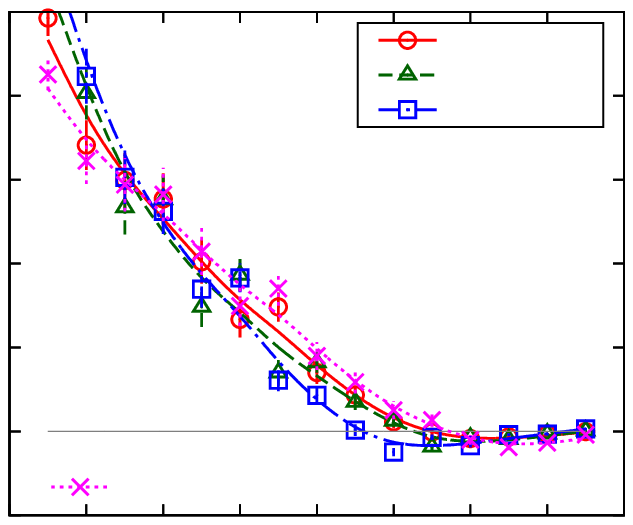}
  \input{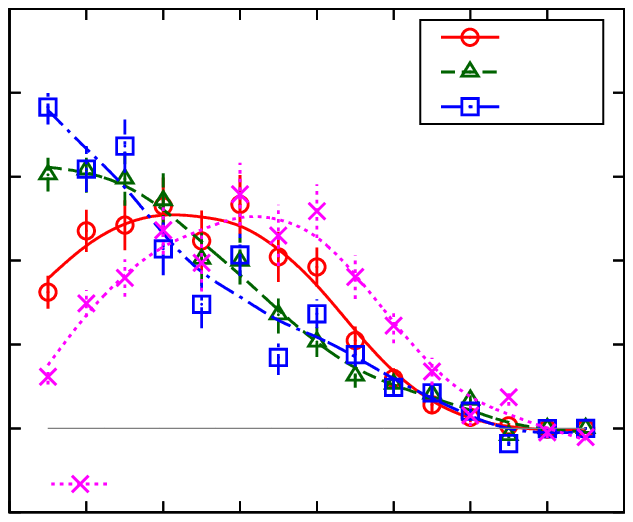}
 \end{center}

 \caption{Optimal weights for 300~$\mu$m silicon (upper left) and 1~cm neon
(upper right). Values are shown for $\beta\gamma =$ 1.00, 3.17 and 10.0.
Optimal weights at $\beta\gamma = 3.17$ are shown for 300, 600 and 1200~$\mu$m
silicon (lower left); 1, 2 and 4~cm neon (lower right).
All results are given for tracks with 15 hits ($i = 1, \dots, 15$).
For comparison the optimal weights of the geometric mean are also
shown (triangles down and crosses).
The lines are drawn to guide the eye.}

 \label{fig:weights}

\end{figure*}

\section{Weighted mean with different path lengths}
\label{sec:pathLength}

In this study we assumed that the path lengths for each hit in the sensitive
detector are the same. In case of real particle trajectories there is a
variance due to bending in the magnetic field, placement of the detector units.
The energy deposits can be corrected towards a reference path length.
The distribution of energy deposit $\Delta$ depends on the
velocity $\beta$ of the particle and the thickness of the traversed material
$t$. To a good approximation, the most probable energy loss $\Delta_p$ and the
full width of the distribution at half maximum $\Gamma_\Delta$ are
\cite{Bichsel:1988if}
\begin{align}
 \Delta_p &= \xi \left[\ln\frac{2 mc^2 \beta^2 \gamma^2 \xi}{I^2}
             + 0.2000 - \beta^2 - \delta \right] \nonumber \\
 \Gamma_\Delta &= 4.018 \xi \label{eq:gamma} \\
\intertext{where}
  \xi &= \frac{K}{2} z^2 \frac{Z}{A} \rho \frac{t}{\beta^2} \nonumber
\end{align}

\noindent is the Landau parameter; $K = 4\pi N_A r_e^2 m_e c^2 =
0.307~075~\mathrm{MeV~cm^2/mol}$; $z$ is the charge of the particle in electron
charge units; $Z$, $A$ and $\rho$ are the mass number, atomic number and the
density of the material, respectively \cite{Nakamura:2010zzi}. Let us consider
the distribution of $y = \Delta/t$ values for a given particle, that is, at a
fix $\beta$. The width of its distribution is independent of $t$
(Eq.~\eqref{eq:gamma}), while the most probable value $y_p$ scales with $t$ as
\begin{equation*}
 y_p(t) = y_p(t_0) +
  \frac{K}{2} z^2 \frac{Z}{A} \rho \frac{\ln(t/t_0)}{\beta^2}
\end{equation*}

\noindent where $t_0$ denotes a fixed reference thickness. The slight $t$
dependence can be minimized by correcting each measurement to $y(t_0)$. For
that, $\beta$ can be estimated from $y$, obtained from the deposits
without the above discussed path length correction.

\begin{figure*}[!h]

 \begin{center}
  \input{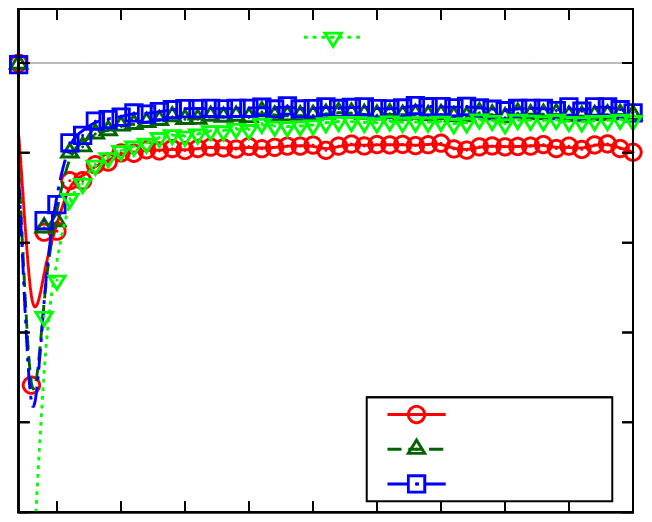}
  \input{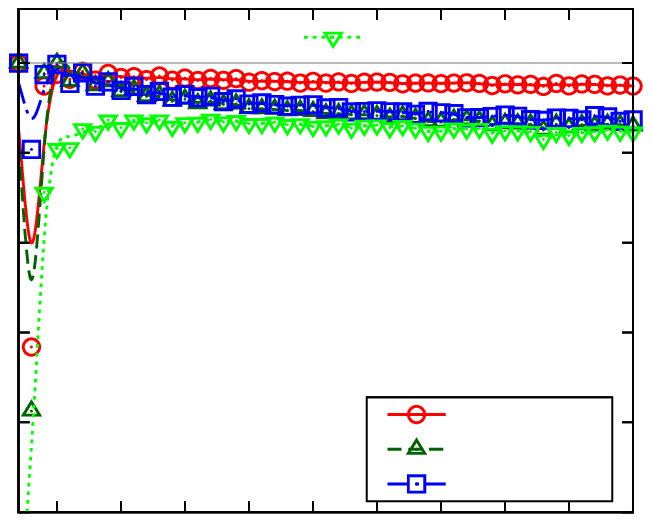}
 \end{center}
 
 \vspace{-0.3in}
 
 \begin{center}
  \input{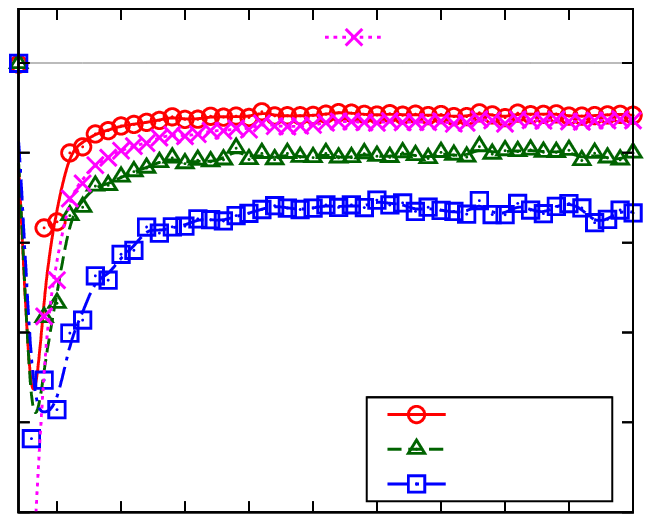}
  \input{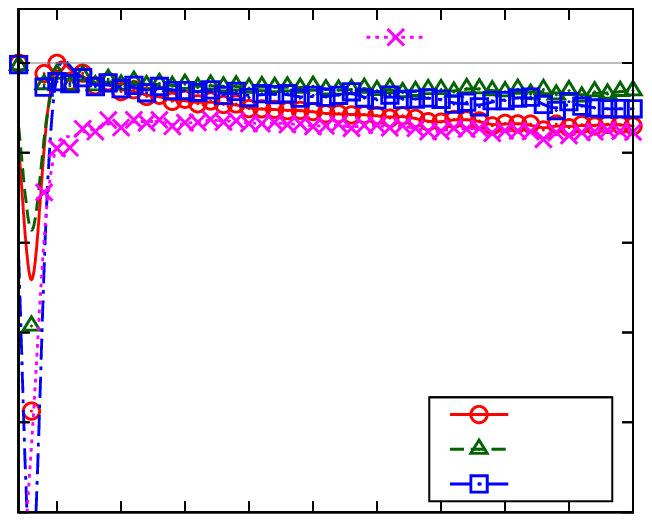}
 \end{center}

 \caption{Performance of the optimized estimator. The ratio of the relative
resolutions (weighted over simple truncated mean) is given as a function of
number of hits on track $n$: for 300~$\mu$m silicon (upper left), and 1~cm neon
(upper right), at various $\beta\gamma$ values; for silicon (lower left), and
neon (lower right), at $\beta\gamma = 3.17$ and various thicknesses.
For comparison the performance of optimized weighted geometric mean is also
shown (triangles down and crosses). Lines are drawn to guide the eye.}

 \label{fig:performance}

\end{figure*}

\section{Results}
\label{sec:results}

Particle identification and yield extraction in the statistical sense are
particularly difficult at those momenta where the differential energy losses of
different type of particles are close. For hadrons, the pion-kaon
resolution gets problematic above about 0.8~GeV/$c$, while for the pion-proton
case it happens above about 1.6~GeV/$c$. Hence the relevant $\beta\gamma$ region is
$1-10$.

In this work charged particles with $\beta\gamma =$ 1.00, 3.16 and 10.0 are
studied, with number of hits $2-50$. Both semiconductor and gaseous detectors
are investigated: for silicon thicknesses of 300, 600 and 1200~$\mu$m,
while for neon with 1, 2 and 4~cm are chosen. For each study one million particles are
used. (In order to speed up computation several million hits in 300~$\mu$m
silicon and 1~cm neon were generated beforehand, for each $\beta\gamma$
settings, and later combined for longer path-length deposits.)

Comparisons of differential energy loss distributions and the estimators obtained
via the simple truncated mean and the weighted arithmetic mean after
optimization are shown in Fig.~\ref{fig:estimators}, in case of 6 hits on
track.

\begin{figure*}[!h]

 \begin{center}
  \input{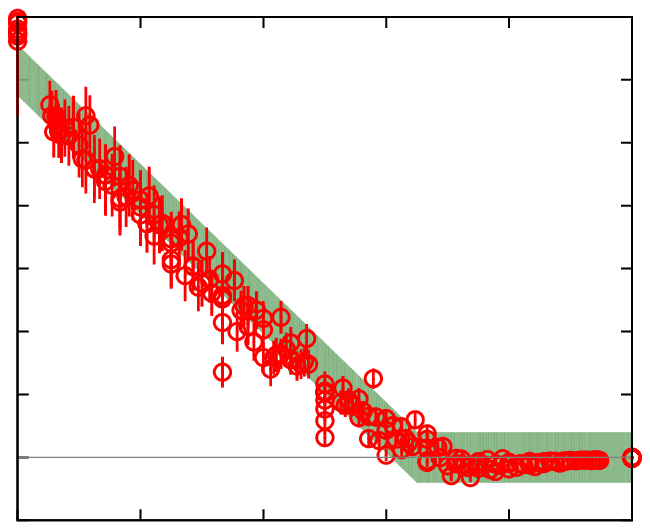}
  \input{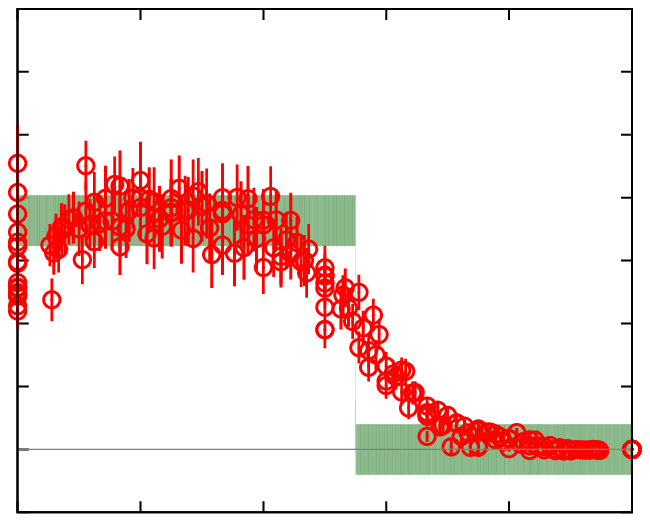}
 \end{center}

 \caption{Optimal weights scaled with the number of hits ($n\cdot w_i$) as a
function of normalized hit number $(i-1)/(n-1)$, if $4 \le n \le 20$, at
$\beta\gamma = 3.17$. Values are shown for 300~$\mu$m silicon (left) and 1~cm
neon (right). The shaded regions indicate a simple description of the weights:
a combination of linear and constant functions with changes at 0.65 for
silicon, and 0.55 for neon.}

 \label{fig:weights_hits}

\end{figure*}

Optimal weights for 300~$\mu$m silicon and 1~cm neon are shown in
Fig.~\ref{fig:weights}-upper, for several $\beta\gamma$ values. Optimal weights
for silicon and neon at $\beta\gamma = 3.17$ are plotted in
Fig.~\ref{fig:weights}-lower, for several thicknesses. All results are given
for tracks with 15 hits, $i = 1, \dots, 15$.
The weights are remarkably independent of $\beta\gamma$ and material thickness for
silicon, while some changes with increasing thickness are seen for neon.  In
case of silicon the hits $10 \le i \le 15$ have very small, in some cases even
negative weights, while the lowest deposits have the highest values.  In case
of 1~cm neon the hits $1 \le i \le 8$ have roughly equal relevance, while the rest
of the hits is not important.
For comparison the optimal weights of the geometric mean are also
shown. While for silicon there is good agreement with arithmetic mean weights,
for neon the numbers somewhat differ but they show similar qualitative
features.

The performance of the optimized estimator can be expressed as the ratio of the
relative resolutions (weighted over simple truncated mean). These are shown as a
function of number of hits on track at various $\beta\gamma$ values
(Fig.~\ref{fig:performance}-upper), and thicknesses
(Fig.~\ref{fig:performance}-lower).
It is clear that there is substantial improvement for both silicon and neon,
for all $\beta\gamma$ and thickness values. In case of few hits (e.g. $n=3$) the
resolution decreased by 20-30\% with respect to the simple truncation. Note
that the improvement tends to a limiting value and the relative ratio is
{\it steadily} below 1 for many hits.
For comparison the performance of the optimized weighted geometric mean is also
plotted. It essentially shows a behavior simular to that of the arithmetic
mean, although with a better improvement at very low $n$ for neon.

\subsection{Other considerations}
\label{sec:other}

Detector and readout noise are more important for silicon since their
magnitude is higher with respect to the energy deposit. Apart from the
standard values (2~keV for silicon, 0.01~keV for neon), noise dependence of
optimal weights was studied (Fig.~\ref{fig:limiting}). While higher
values do not influence the weights for neon, in case of silicon at 10~keV
their distribution starts to deform into a box distribution.  This behavior
can be understood: the addition of the Gaussian noise softens the lower leading
edge of the energy deposit distribution (Fig.~\ref{fig:distro}-left) and makes
the lower values less important.

In a more complete detector simulation the effects of readout threshold
(underflow, left truncation) and the upper limit of detector linearity (right
censoring, overflow) should be taken into account. Still, this latter is likely
not important since the highest deposits in any case will have low weights.

\subsection{Universality, connections}

While the list of weights as function of number of hits can be tabulated
(see App.~\ref{sec:few}), it would be much easier to find a simpler description.
Optimal weights scaled with the number of hits ($n\cdot w_i$) as a function of
normalized hit number $z \equiv (i-1)/(n-1)$ are plotted in
Fig.~\ref{fig:weights_hits}, if $4 \le n \le 20$. The dependence on the
normalized hit number can be easily described as a combination of linear and
constant functions. For silicon, with $z_\text{Si} \approx 0.65$,
\begin{align}
 \label{eq:silicon}
 n\cdot w_i &= \begin{cases}
                2(z_\text{Si} - z)/z_\text{Si}^2 & \text{if $z < z_\text{Si}$} \\
                0, & \text{otherwise}
               \end{cases} \\
\intertext{while for neon, with $z_\text{Ne} \approx 0.55$,}
 \label{eq:neon}
 n\cdot w_i &= \begin{cases}
                1/z_\text{Ne} & \text{if $z < z_\text{Ne}$} \\
                0, & \text{otherwise}.
               \end{cases}
\end{align}

In case of many measurements, the optimal weights can be obtained using the
energy deposit distribution, App.~\ref{sec:many} gives a detailed
derivation.

The simple functional forms described in Eqs.~\eqref{eq:silicon} and
\eqref{eq:neon} have a deeper cause, namely it is strongly connected to the
functional form of the deposit distribution. For detailed argumentation see
App.~\ref{sec:equally}. If the density (Fig.~\ref{fig:distro}) can be
locally described by
\begin{itemize}

 \item a power function, the local weights are zero;

 \item a product of exponential and power functions, the local weights
are constant;

 \item a product of Gaussian and power functions, the local weights are linear in $z$.

\end{itemize}

This study was restricted to linear combination of measurements. It can be
shown that while for semiconductor detectors the optimized weighted mean
estimator may be further improved by using maximum likelihood methods, for
gaseous detectors the simple (0\%,55\%) truncation (Eq.~\eqref{eq:neon})
already gives excellent results (see App.~\ref{sec:connection}).

\section{Conclusions}

The estimation of differential energy loss for charged particles in tracker
detectors was studied. It was shown that the simple truncated mean method can
be generalized to the linear combination of the energy deposit measurements.
The optimized weights are rather independent of particle momentum and material
thickness, allowing for a robust estimation. Weighted arithmetic and geometric
means result in better particle separation power for both semiconductor and
gaseous detectors. Further inspections showed that weights are deeply
connected to corresponding energy deposit distribution, allowing for a simple
universal description of weights as function of number of hits.

\section*{Acknowledgements}

The authors wish to thank to Kriszti\'an Krajcz\'ar for helpful discussions.
This work was supported by the Hungarian Scientific Research Fund with the
National Office for Research and Technology (K 48898, K 81614, H07-B 74296),
and the CERN Summer Student Program.

\appendix

\begin{table*}[!h]

 \caption{Optimal weights scaled with the number of hits ($n\cdot w_i$) for 300
$\mu$m silicon, at $\beta\gamma = 3.17$, in case if hit numbers $n = 2, \dots,
9$. Errors indicate the sensitivity corresponding to 1\% increase of the
relative resolution of the estimator.}

 \label{tab:si}

 \begin{center}
  \begin{tabular}{lccccccccc}
\hline
 $i$ & $n=2$ & $n=3$ & $n=4$ & $n=5$ & $n=6$ & $n=7$ & $n=8$ & $n=9$ \\
\hline
 $1$  & 2.0 $\pm$ 3.6 & 3.0 $\pm$ 2.8 & 3.4 $\pm$ 0.7 & 3.3 $\pm$ 0.4 & 3.3 $\pm$ 0.3 & 3.4 $\pm$ 0.3 & 3.4 $\pm$ 0.2 & 3.5 $\pm$ 0.2 \\

 $2$  & -0.0 $\pm$ 0.1 & 0.0 $\pm$ 0.1 & 0.7 $\pm$ 0.1 & 1.5 $\pm$ 0.2 & 1.9 $\pm$ 0.2 & 2.0 $\pm$ 0.3 & 2.2 $\pm$ 0.3 & 2.3 $\pm$ 0.3 \\

 $3$  & --  & -0.0 $\pm$ 0.1 & -0.0 $\pm$ 0.1 & 0.2 $\pm$ 0.1 & 0.8 $\pm$ 0.1 & 1.3 $\pm$ 0.2 & 1.5 $\pm$ 0.2 & 1.7 $\pm$ 0.2 \\

 $4$  & --  & --  & -0.0 $\pm$ 0.1 & -0.0 $\pm$ 0.1 & 0.0 $\pm$ 0.1 & 0.4 $\pm$ 0.1 & 0.8 $\pm$ 0.1 & 1.0 $\pm$ 0.1 \\

 $5$  & --  & --  & --  & -0.0 $\pm$ 0.1 & -0.0 $\pm$ 0.1 & -0.0 $\pm$ 0.1 & 0.1 $\pm$ 0.1 & 0.5 $\pm$ 0.1 \\

 $6$  & --  & --  & --  & --  & -0.0 $\pm$ 0.1 & -0.0 $\pm$ 0.1 & -0.1 $\pm$ 0.1 & 0.1 $\pm$ 0.1 \\

 $7$  & --  & --  & --  & --  & --  & -0.0 $\pm$ 0.1 & -0.0 $\pm$ 0.1 & -0.1 $\pm$ 0.1 \\

 $8$  & --  & --  & --  & --  & --  & --  & -0.0 $\pm$ 0.1 & -0.0 $\pm$ 0.1 \\

 $9$  & --  & --  & --  & --  & --  & --  & --  & -0.0 $\pm$ 0.1 \\

\hline
\end{tabular}

 \end{center}

\end{table*}

\begin{table*}[!h]

 \caption{Optimal weights scaled with the number of hits ($n\cdot w_i$) for 1
cm neon, at $\beta\gamma = 3.17$, in case if hit numbers $n = 2, \dots, 9$.
Errors indicate the sensitivity corresponding to 1\% increase of the relative
resolution of the estimator.}

 \label{tab:ne}

 \begin{center}
  \begin{tabular}{lccccccccc}
\hline
 $i$ & $n=2$ & $n=3$ & $n=4$ & $n=5$ & $n=6$ & $n=7$ & $n=8$ & $n=9$ \\
\hline
 $1$  & 2.0 $\pm$ 9.4 & 2.8 $\pm$ 0.8 & 2.3 $\pm$ 0.3 & 2.0 $\pm$ 0.2 & 1.9 $\pm$ 0.2 & 1.7 $\pm$ 0.2 & 1.6 $\pm$ 0.2 & 1.6 $\pm$ 0.2 \\

 $2$  & -0.0 $\pm$ 0.1 & 0.2 $\pm$ 0.1 & 1.6 $\pm$ 0.2 & 2.0 $\pm$ 0.3 & 2.1 $\pm$ 0.3 & 2.1 $\pm$ 0.3 & 2.0 $\pm$ 0.3 & 2.0 $\pm$ 0.2 \\

 $3$  & --  & -0.0 $\pm$ 0.1 & 0.1 $\pm$ 0.1 & 1.0 $\pm$ 0.1 & 1.4 $\pm$ 0.2 & 1.9 $\pm$ 0.3 & 2.0 $\pm$ 0.3 & 1.9 $\pm$ 0.3 \\

 $4$  & --  & --  & -0.0 $\pm$ 0.1 & 0.0 $\pm$ 0.1 & 0.5 $\pm$ 0.1 & 1.0 $\pm$ 0.1 & 1.5 $\pm$ 0.2 & 1.8 $\pm$ 0.2 \\

 $5$  & --  & --  & --  & -0.0 $\pm$ 0.1 & 0.0 $\pm$ 0.1 & 0.3 $\pm$ 0.1 & 0.7 $\pm$ 0.1 & 1.1 $\pm$ 0.2 \\

 $6$  & --  & --  & --  & --  & -0.0 $\pm$ 0.1 & -0.0 $\pm$ 0.1 & 0.2 $\pm$ 0.1 & 0.5 $\pm$ 0.1 \\

 $7$  & --  & --  & --  & --  & --  & -0.0 $\pm$ 0.1 & -0.0 $\pm$ 0.1 & 0.1 $\pm$ 0.1 \\

 $8$  & --  & --  & --  & --  & --  & --  & -0.0 $\pm$ 0.1 & -0.0 $\pm$ 0.1 \\

 $9$  & --  & --  & --  & --  & --  & --  & --  & -0.0 $\pm$ 0.1 \\

\hline
\end{tabular}

 \end{center}

\end{table*}

\section{Optimal weights in case of few measurements}
\label{sec:few}

The obtained weights for 300 $\mu$m silicon and 1 cm neon, at $\beta\gamma =
3.17$, in case of hit numbers $2 \le n \le 9$ are shown in Tables~\ref{tab:si}
and \ref{tab:ne}, respectively. Errors indicate the sensitivity corresponding
to 1\% increase of the relative resolution of the estimator.

\section{Optimal weights in case of many measurements}
\label{sec:many}

A random variable $x$ is described by the probability density function $f(x)$
and cumulative distribution function $F(x) = \int_{-\infty}^x f(x') dx'$.
During an observation $f$ is sampled $n$ times, these measurements are
rearranged to increasing order ($x_i \le x_{i+1}$, $i = 1,2,\dots,n$). We are
interested in the means $m_i$ and covariance $V_{ij}$ of the ordered samples in
the continuous limit ($n \gg 1$).

\subsection{Means and covariance of measurements}
\label{sec:many_calc}

The probability density that $x_i$ is the $i$th measurement
\begin{equation*}
 p(x_i) = \frac{n!}{(i-1)! (n-i)!} F(x_i)^{i-1} f(x_i) [1 - F(x_i)]^{n-i}.
\end{equation*}

\noindent
It is easier to work with $-\log p$, since its minimum gives the most probable
value $\overline{x_i}$. In the Gaussian approximation the mean $m_i =
\overline{x_i}$ and its variance $V_{ii}$ is also calculable.  If $1 < i < n$,
then the factors containing $F$ already constrain well enough the position of
$m_i$, hence $f(x_i)$ can be approximated by a constant:
\begin{equation}
 p(x_i) \approx \frac{n!}{(i-1)! (n-i)!} F(x_i)^{i-1} [1 - F(x_i)]^{n-i}.
 \label{eq:1dProbDensity}
\end{equation}
 
\noindent
At $x_i = m_i$
\begin{align*}
 [-\log p(x)]' &= 0, &
 1/V_{ii}  &= [-\log p(x)]''.
\end{align*}

\noindent
By solving the equation on the left, we get
\begin{align}
 F(m_i) &= \frac{i-1}{n-1}, &
 V_{ii}            &= \frac{(i-1)(n-i)}{(n-1)^3 f^2(m_i)}.
 \label{eq:variance}
\end{align}

\noindent Note again that the above approximations are valid only for $1<i<n$.
(For the lowest and highest measurements we would get $m_1 =
-\infty$, $m_n = \infty$, and $\sigma_1^2 = \sigma_n^2 = \infty$.)

The probability density that $x_i$ and $x_j$ are the $i$th and $j$th
measurements ($i \le j$), respectively,
\begin{multline*}
 p(x_i,x_j) = \frac{n!}{(i-1)! (j-i-1)! (n-j)!} \cdot \\ \cdot
              F(x_i)^{i-1} [F(x_j) - F(x_i)]^{j-i-1}
              [1 - F(x_j)]^{n-j}.
\end{multline*}

\noindent Similarly to Eq.~\eqref{eq:1dProbDensity}, by assuming $1<i<j<n$, we
have omitted the factors $f(x_i)$ and $f(x_j)$.

Close to its minimum $p(x_i,x_j)$ can be approximated by a multivariate normal
distribution.
At the minimum ($x_i = m_i$ and $x_j = m_j$) the correlation coefficient is
\begin{equation*}
 \rho_{ij} =
       {      \frac{\p^2 \log p}{\p x_i \p x_j}} \; \Biggl/
       {\sqrt{\frac{\p^2 \log p}{\p x_i^2}
              \frac{\p^2 \log p}{\p x_j^2}}}.
\end{equation*}

\noindent
The partial derivatives are
\begin{align*}
 \frac{\p^2 \log p}{\p x_i \p x_j} &=
  (n-1)^2 \frac{j-i-1}{(j-i)^2} f(m_i) f(m_j)) \\
 \begin{split}
 \frac{\p^2 \log p}{\p x_i^2}     &= 
  \frac{n-1}{j-i} f'(m_i) - \\ &-
   (n-1)^2 \frac{(j-i)(j-1) - (i-1)}{(i-1) (j-i)^2} f^2(m_i)
 \end{split} \\ 
 \begin{split}
 \frac{\p^2 \log p}{\p x_j^2}     &=
  -\frac{n-1}{j-i} f'(m_j) - \\ &-
   (n-1)^2 \frac{(j-i)(n-i) - (n-j)}{(n-j) (j-i)^2} f^2(m_j).
 \end{split}
\end{align*}

In the $n \gg 1$ limit the coefficients of $f'$ can be neglected
if compared to the ones of the $f^2$ terms. Similarly the nominators of the
coefficient of $ff$ in all three cases can be simplified by neglecting the
terms $1$, $(i-1)$ and $(n-j)$, respectively:
\begin{align*}
 \frac{\p^2 \log p}{\p x_i \p x_j}  &\approx
  (n-1)^2 \frac{1}{j-i} f(m_i) f(m_j) \\
 -\frac{\p^2 \log p}{\p x_i^2}      &\approx
   (n-1)^2 \frac{(j-1)}{(i-1) (j-i)} f^2(m_i) \\
 -\frac{\p^2 \log p}{\p x_j^2}      &\approx
   (n-1)^2 \frac{(n-i)}{(n-j) (j-i)} f^2(m_j).
\end{align*}

\noindent
With that the correlation coefficient is
\begin{equation*}
 \rho_{ij} = \sqrt{\frac{(i-1) (n-j)}{(n-i) (j-1)}}.
\end{equation*}

\noindent Note that $\rho_{ij}$ does not depend on $f$. The covariance of the
$i$th and $j$th measurements ($i \le j$) is
\begin{equation}
 V_{ij} = \rho_{ij} \sqrt{V_{ii} V{jj}} = 
  \frac{(i-1) (n-j)}{(n-1)^3 f(m_i) f(m_j)}.
 \label{eq:covariance}
\end{equation}

\noindent
The expression gives nicely back the variance ($i=j$) already obtained in
Eq.~\eqref{eq:variance}-right.

\subsection{Optimal weights} 
\label{sec:many_optimal}

With the vector of means $m_i$ (Eq.~\eqref{eq:variance}-left) and the
covariance matrix $V_{ij}$ (Eq.~\eqref{eq:covariance}), the optimal weights are
calculable. 
In the high $n$ limit, we can consider the following continuous variables
\begin{align*}
 a &\equiv \frac{i-1}{n-1}, & b &\equiv \frac{j-1}{n-1}, & 0<a,b<1.
\end{align*}

\noindent
The means and covariance in this variables are
\begin{align*}
 m(a) &= F^{-1}(a) \\
 V(a,b) & = \operatorname{min}(a,b) [1 - \operatorname{max}(a,b)] m'(a) m'(b).
\end{align*}

On the analogy of $m = Vw$ 
there is a
Friedholm integral equation of the first kind for $w$
\begin{multline*}
 m(a) = \int_0^1 V(a,b) w(b) \d b = \\ = 
        \int_0^1 \operatorname{min}(a,b)
                 [1 - \operatorname{max}(a,b)] m'(a) m'(b) w(b) \d b.
\end{multline*}

\noindent
It can be written in the form
\begin{gather*}
 g(a) = \int_0^1 K(a,b) y(b) \d b
\intertext{where}
 g(a) = \frac{m(a)}{m'(a)}, \qquad y(b) = m'(b) w(b) \\
 K(a,b) = \operatorname{min}(a,b) [1 - \operatorname{max}(a,b)].
\end{gather*}

\noindent
The kernel $K$ is a special one
\begin{gather*}
 g(a) = \int_0^a (1-a) b y(b) \d b + \int_a^1 a (1-b) y(b) \d b \\
 g'(a) = - \int_0^a b y(b) \d b + \int_a^1 (1-b) y(b) \d b \\
\intertext{The equation can be solved by two consecutive derivations in $a$
giving the result}
 g''(a) = - y(a).
\end{gather*}

In summary the optimal weights $w$ in the continuous variable $a$ are
\begin{equation}
 w = - \frac{1}{m'} \left(\frac{m}{m'}\right)''.
 \label{eq:w}
\end{equation}

\noindent With the energy loss $m$ as variable, for a weight at $F(m)$

\begin{multline}
 w[F(m)] = - \frac{f'}{f} + m \left(\frac{f'}{f}\right)^2 - m \frac{f''}{f}
 = \\ = - \left[m (\log f)'\right]'.
 \label{eq:optimal}
\end{multline}

\begin{figure*}

 \begin{center}
  \input{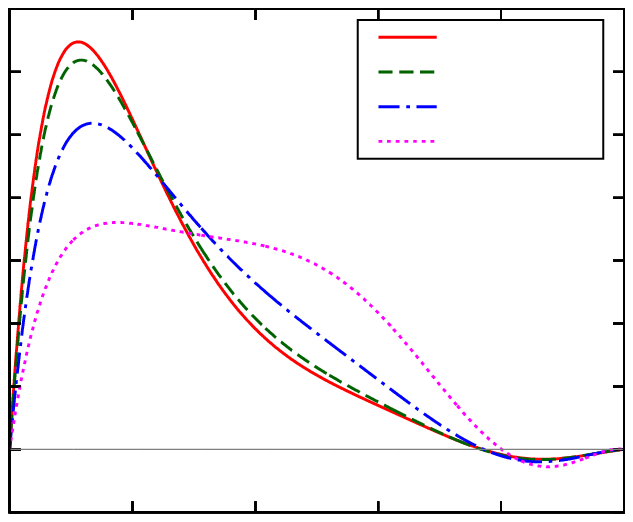}
  \input{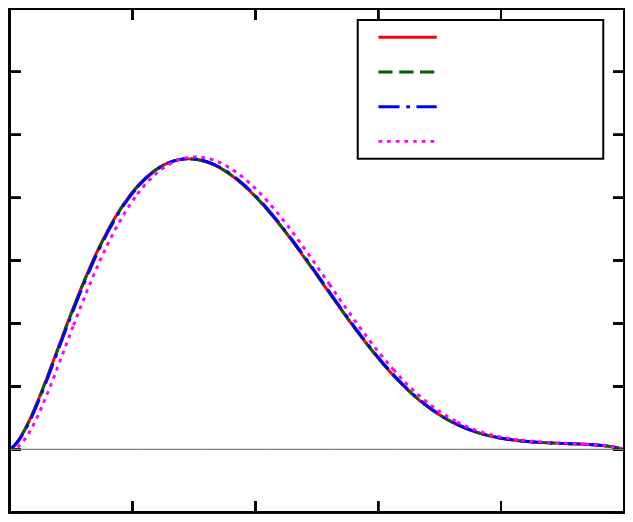}
 \end{center}

 \caption{Optimal weights scaled with the number of hits ($n\cdot w$) as a
function of normalized hit number $(i-1)/(n-1)$, if $n$ is big, at $\beta\gamma
= 3.17$. Values are shown for 300~$\mu$m silicon (left) and 1~cm neon (right),
both with several noise settings.}

 \label{fig:limiting}

\end{figure*}

With help of this exact result (Eq.~\eqref{eq:optimal}),
the optimal weights scaled with the number of hits ($n\cdot w_i$) as a
function of normalized hit number $z \equiv (i-1)/(n-1)$ are shown in
Fig.~\ref{fig:limiting}, at $\beta\gamma = 3.17$, if $n$ is big.

The minimum of the relative resolution of the weighted average is
\begin{equation*}
  \frac{1}{\sqrt{\int_0^1 m w}}
  = \frac{1}{\sqrt{- \int_0^1 \frac{m}{m'} \left(\frac{m}{m'}\right)''}}.
\end{equation*}

\section{Irrelevant and equally relevant measurements} 
\label{sec:equally}

In the continuous limit we can find probability distribution functions
$f$ with special characteristics, by solving Eq.~\eqref{eq:w}.
 
The ordered measurements are {\it irrelevant} if $w=0$. The condition $1/m' = 0$
would give $f = 0$, a meaningless unphysical solution. On the other hand if
\begin{equation*}
 \left(\frac{m}{m'}\right)'' = 0 \qquad 
  \Rightarrow \qquad m(z) = \left(\frac{z + d/(b+1)}{a}\right)^{1/(b+1)}
\end{equation*}
\begin{equation}
 f(m) = \frac{1}{m'(z)} = a \cdot m^b.
 \label{eq:w=0}
\end{equation}

\noindent
Hence measurements from distributions with power functions should be
neglected, their optimal weights are zero.

The ordered measurements are {\it equally relevant} if
\begin{gather*}
 w = - \frac{1}{m'} \left(\frac{m}{m'}\right)'' = c = const > 0. \\
\intertext{The equation}
 - f[m(z)] \frac{\p^2 \left(m(z) f[m(z)]\right)}{\p z^2} = c \\
\intertext{can be solved by resolving the composition $f \circ m$, giving}
 f(m) = a \cdot m^b \exp(-c \cdot m),
\end{gather*}

\noindent
which is the product of a power and an exponential function (compare with
Fig.~\ref{fig:distro}-right and Fig.~\ref{fig:weights_hits}-left, the case of
neon).  Note that $c=0$ indeed gives back the result in Eq.~\eqref{eq:w=0}.  It
can be shown that weights are a linear function of $z$ if $f(m)$ is the product
of a power and a Gaussian function (compare with Fig.~\ref{fig:distro}-left and
Fig.~\ref{fig:weights_hits}-left, the case of silicon).

The distribution of energy loss $f(\Delta)$ examined in this study has a
$1/\Delta^2$ behavior for large energy deposits due to Coulomb excitations.
(The Landau distribution, which is often used to model and approximate energy
loss, also has a $1/\Delta^2$ power-law tail.) This is why we got small optimal
weights for the upper measurements. It is also the fundamental cause for the
success of the classical truncated mean method.

\section{Connection to maximum likelihood estimation}
\label{sec:connection}

In this work the optimization of differential energy loss estimation was
confined to the linear combination of the measurements, or of a monotonic
function of the measurements. Are there cases when this relatively simple
prescription is close to the performance of a, supposedly more powerful,
maximum likelihood estimation?

Let us assume that the scale of energy loss distribution is characterized by
the most probable deposit $y_P$, such that the probability of a deposit $y_i$
is given by 
\begin{gather*}
 \frac{y_0}{y_P} f\left(y_i\frac{y_0}{y_P}\right)
\intertext{where $f$ is a universal function and $y_0$ is constant. In that case the likelihood associated to a
track with $n$ hits is}
 P = \prod_{i=1}^n \frac{y_0}{y_P} f\left(y_i\frac{y_0}{y_P}\right)
\intertext{and}
 -\log P = \sum_{i=1}^n \left[\log \frac{y_P}{y_0} - \log
f\left(y_i \frac{y_0}{y_P}\right)\right].
\intertext{Its value is extremum if the derivative is zero, giving}
 y_p = y_0 \sum_{i=1}^n \frac{y_i}{n} \cdot (-\log f)' \left(y_i
\frac{y_0}{y_P}\right).
\end{gather*}

\noindent
The most probable value $y_P$ can be obtained by a simple linear
combination of measurements if, for an interval around $y_i$, the function
$(-\log f)'$ behaves as
\begin{gather*}
 (-\log f)' (y) = c_i - b_i/y \\
\intertext{where $c_i$ and $b_i$ are local constants. With this}
 y_P = \frac{y_0}{\sum_{i=1}^n (1+b_i)} \sum_{i=1}^n c_i y_i.
\intertext{The corresponding functional form is}
 f(y) = a \cdot y^b \exp(-c \cdot y)
\end{gather*}

\noindent that exactly matches the form found for neon (see
Fig.~\ref{fig:distro}-right and App.~\ref{sec:equally}).

In summary we can say that while for semiconductor detectors the optimized
weighted mean estimator may be further improved with maximum likelihood
methods, for gaseous detectors the simple (0\%,55\%) truncation already gives
excellent results.

\bibliographystyle{elsarticle-num}
\bibliography{weightedMean}

\end{document}